\documentclass[useAMS,usenatbib]{mn2e}

\usepackage{amssymb}
\usepackage{amsfonts}
\usepackage{times}
\usepackage{mncite}
\usepackage{epsfig}
\usepackage{psfig}
\usepackage{verbatim}

\newcommand{\msun}{M_{\sun}}

\begin{document}

\title[A self-gravitating disc in L1527 IRS?]
{A self-gravitating disc around L1527 IRS?}

\author[Duncan Forgan \& Ken Rice]
 {Duncan Forgan$^1$\thanks{E-mail: dhf@roe.ac.uk}, Ken Rice$^1$ \\
$^1$ SUPA\thanks{Scottish Universities Physics Alliance},
Institute for Astronomy, University of Edinburgh, Blackford Hill, Edinburgh, EH9 3HJ \\}

\maketitle
\begin{abstract}

Recent observations of the Class 0 protostar L1527 IRS have revealed a
rotationally supported disc with an outer radius of at least 100 au.
Measurements of the integrated flux at 870 $\mu$m suggest a disc mass
that is too low for gravitational instability to govern angular
momentum transport. However, if parts of the
disc are optically thick at sub-mm wavelengths, the sub-mm fluxes will
underestimate the disc mass, and the disc's actual mass may be
substantially larger, potentially sufficient to be self-gravitating.

We investigate this possibility using simple self-gravitating disc
models.  To match the observed mass accretion rates requires a
disc-to-star mass ratio of at least $\sim 0.5$, which produces sub-mm fluxes that
are similar to those observed for L1527 IRS in the absence of irradiation from the envelope or central star.  If irradiation is significant, then the predicted fluxes exceed the observed fluxes by around an order of magnitude.  Our model also indicates
that the stresses produced by the gravitational instability are low enough to prevent disc fragmentation.

As such, we conclude that observations do not rule out the possibility
that the disc around L1527 IRS is self-gravitating, but it is more likely that despite being a very young system, this disc may already have left the self-gravitating phase.

\end{abstract}

\section{Introduction}
It is now generally accepted that low-mass stars form through the
collapse of cold, dense molecular cloud cores \citep{terebey84}. These
cores, however, typically have rotation rates that mean that all
the mass cannot fall directly onto the young protostar
\citep{caselli02}. A mechanism is therefore required to transport this excess
angular momentum away, allowing mass accretion to take place. 

The standard scenario is that a large fraction of the material will
fall onto a disc around the young protostar, and that this disc will
provide the mechanism for transporting the angular momentum outwards.
In many astrophysical discs, this angular momentum transport could be
driven by magnetohydrodynamic (MHD) turbulence initiated by the
magnetorotational instability (MRI; \citealt{balbus91}). At very early
times, these cold, dense molecular cloud cores are, however, unlikely
to be sufficiently ionized to sustain MHD turbulence \citep{blaes94}.
Instead, it is thought that these discs could be massive and that disc
self-gravity could provide the dominant transport mechanism during the
earliest stages of star formation \citep{toomre64,lin86,laughlin94}.

Although discs around very young (class 0 and class I) protostars
could be massive, they are difficult to identify because they are
still heavily embedded in their nascent molecular cloud cores.
Typically, high-resolution interferometic observations at sub-mm and
mm wavelengths are used to resolve out the surrounding envelope
\citep{rodriguez05,eisner05}.  In some cases these discs do appear to
be quite massive (see, for example, \citet{greaves11} and references
therein). Interferometric observations by \citet{maury10} at 1.3 mm,
however, suggest that discs around very young protostars are very
compact ($< 30$ AU). Magnetohydrodynamic simulations of collapsing
molecular cloud cores \citep{hennebelle08} suggest that even moderate
magnetic fields can strongly influence angular momentum transport, and can
prevent the disc becoming extended at these early times.  This may
imply that disc self-gravity is not particularly important at this stage.  

By contrast, recent observations of the protostar L1527 IRS
\citep{tobin12} have revealed an edge-on disc with an outer radius of
at least $100$ AU.  Since this disc is edge-on, they were able to
determine the rotational velocity of the material in the disc and
could confirm that it was indeed in Keplerian rotation about a central
protostar, with a mass of $\sim 0.2$ M$_\odot$.  Not only is this the
first confirmation of a disc around a class 0 protostar, it also
indicates that these discs can be reasonably extended ($> 100$
AU). Observations suggest that most of the luminosity is due to
accretion through the disc, giving an accretion rate of $6.6 \times
10^{-7}$ M$_\odot$ yr$^{-1}$.  Continuum observations at 870 $\mu$m
indicate a disc mass of $\sim 0.007 \pm 0.0007$ M$_\odot$, about 30 times smaller
than the mass of the central protostar.

If the disc mass estimate is correct, there is insufficient mass for
this disc to be self-gravitating. Angular momentum transport could
then be driven by MRI.  MHD simulations \citep{papaloizou03} suggest
that MRI can produce stresses resulting in a viscous $\alpha$
\citep{shakura73} of up to $\sim \times 10^{-2}$. The steady accretion
rate in a viscous accretion disc with a reasonably low mass and with a
viscous $\alpha$ of this magnitude would typically be $
10^{-7}$ M$_\odot$ yr$^{-1}$ \citep{armitage03}.  This leaves two possibilities:

\begin{enumerate}
\item This disc maintains a high accretion rate due to vigorous MRI activity supplying angular momentum transport, or
\item The disc maintains a high accretion rate due to the gravitational instability, and the disc mass is higher than observations indicate.
\end{enumerate}

There is both theoretical and observational
evidence \citep{vorobyov09,hartmann08} that disc masses around young
protostars have been systematically underestimated, so option ii) may not be
particularly surprising.  If the disc mass is indeed higher than
estimated, disc self-gravity could then become a viable mechanism for
transporting angular momentum.  As self-gravitating discs tend to be more centrally condensed than their non-self-gravitating counterparts \citep{clarke09,rice09}, this provides a satisfying explanation for their ability to hide mass at high optical depths in the hotter inner disc regions.

In this paper we determine the properties of a self-gravitating disc
that could provide the observed accretion rate of L1527 IRS, and compare the
observational signatures of such a disc with
those observed.

\section{Model}
Here we use the model developed by \citet{clarke09} (see also
\citet{rice09}).  A disc is susceptible to the gravitational
instability if the $Q$ parameter \citep{toomre64} is close to, but larger
than, unity. Here we assume that at all radii in the disc
\begin{equation}
Q = \frac{c_s \Omega}{\pi G \Sigma} = 2,
\label{eq:Q}
\end{equation}   
where $c_s$ is the sound speed, $\Omega$ is the angular frequency, $G$
is gravitational constant, and $\Sigma$ is the disc surface density.
It is now known that the angular momentum transport in a quasi-steady
self-gravitating disc can be regarded as viscous-like
\citep{balbus99,lodato04} such that the stresses produce an effective
viscous $\alpha$.  The mass accretion rate, $\dot{M}$, in such a
quasi-steady disc will satisfy
\begin{equation}
\dot{M} = \frac{3 \pi \alpha c_s^2 \Sigma}{\Omega} = {\rm constant}.
\label{eq:Mdot}
\end{equation}
The local effective gravitational $\alpha$ is determined by assuming the disc is in thermal equilibrium \citep{gammie01}, giving
\begin{equation}
\alpha = \frac{4}{9 \gamma (\gamma - 1) t_{\rm cool} \Omega}.
\label{eq:alpha}
\end{equation}
Using Rosseland mean opacities, $\kappa$, from \citet{bell94} the optical depth can be estimated as $\tau = \kappa \Sigma$ and the cooling rate
is then \citep{hubeny90}
\begin{equation}
\Lambda = \frac{8 \sigma T^4}{3 \tau},
\label{eq:coolrate}
\end{equation}
where $T$ is the midplane temperature and $\sigma$ is the Stefan-Boltzmann constant.  The cooling time is then the thermal energy per unit
area ($c_s^2 \Sigma/\gamma(\gamma-1)$) divided by this cooling rate. This is then sufficient
to determine - given $\dot{M}$ and stellar mass, $M_*$ - the sound speed and surface density at any radius in a 
quasi-steady, self-gravitating disc.

If the disc is subject to irradiation, we modify equation (\ref{eq:coolrate}) as follows:

\begin{equation} 
\Lambda = \frac{8 \sigma (T^4 - T_{\rm irr}^4)}{3 \tau},
\end{equation}

\noindent where $T_{\rm irr}$ represents the temperature of the local radiation field.

\section{Results}
The rotational velocity of the material in the disc around L1527 IRS suggest a central protostar mass of $\sim 0.2$ M$_\odot$ 
\citep{tobin12}.  We use the above model to determine the properties of quasi-steady, self-gravitating discs around a central protostar of mass $0.2$ M$_\odot$.  We consider two cases: in the first case, there is no significant external irradiation; in the second, we model the irradiation field from the envelope and central star as $T_{\rm irr} =30$ K, in keeping with the dust temperature estimates of \citet{tobin12}.

\subsection{No Irradiation}
\subsubsection{Basic disc properties}
\begin{figure}
\begin{center}
\psfig{figure=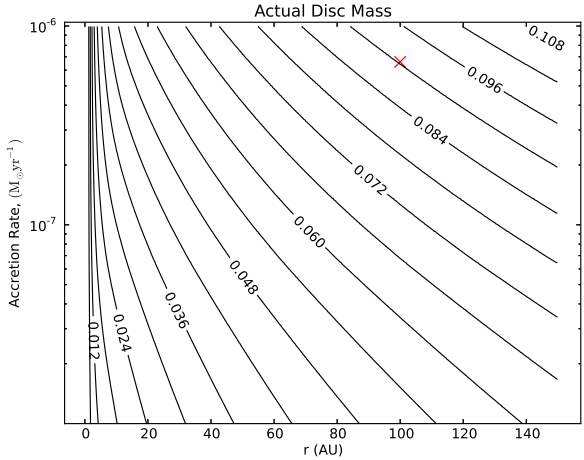,width=0.5\textwidth}
\caption{2D contours of the actual disc mass calculated from the
    self-gravitating disc models, as a function of the
    steady state accretion rate $\dot{M}$, and the disc's outer radius
    $r_{\rm out}$, for a star of mass $0.2 \msun$.  The asterisk indicates
    the required disc mass, assuming an outer radius of 100 au, to produce
a mass accretion rate like that measured for L1527 IRS.}
\label{fig:mdot_r_q}
\end{center}
\end{figure}

As discussed in Section 2, given $\dot{M}$ and $M_*$, we can determine the surface density and temperature at any radius in a quasi-steady, 
self-gravitating disc. Figure \ref{fig:mdot_r_q} shows contours of disc mass (in Solar masses) plotted against accretion rate and outer disc radius.  
The observed accretion rate in L1527 IRS is $\sim 6.6 \times 10^{-7}$ M$_\odot$ yr$^{-1}$ which, if the outer radius is $100$ au, would 
require a disc with a mass of $0.091$ M$_\odot$.  This is shown by the red cross in Figure \ref{fig:mdot_r_q}.  If the outer
radius is $150$ au, this rises to $\sim 0.11$ M$_\odot$.  This is about half the mass of the central star and more than 10 times greater
than that estimated from the observed $870 \mu$m flux from L1527 IRS.

\subsubsection{Mass estimates}
We can use our models to determine the $870 \mu$m flux for our simulated discs.  For a system at a distance $D$, the flux at 
frequency $\nu$ from an annulus of the disc at radius $r$ from the central star, and with radial extent $dr$ is
\begin{equation}
F_\nu(r)dr = \frac{2 k}{c^2 D^2} \nu^2 \kappa(\nu) \Sigma(r) T(r) 2 \pi r dr,
\label{eq:flux}
\end{equation}
where $k$ is Boltzmann's constant, $c$ is the speed of light, and $T(r)$ is the disc midplane temperature.  The frequency 
dependent opacity $\kappa(\nu)$ is
\begin{equation}
\kappa (\nu) = \kappa_o  \left( \frac{\nu}{\nu_o}\right)^\beta
= \kappa_o \left( \frac{\lambda_o}{\lambda}\right)^\beta,
\label{eq:opacity}
\end{equation}
where $\beta$ is typically 1, and $\kappa_o$ is the opacity at a reference wavelength of $\lambda_o = 850 \mu$m.

Quasi-steady, self-gravitating discs are, however, centrally condensed \citep{clarke09,rice09} and it is expected that they
may be optically thick even at mm wavelengths \citep{greaves10}.  We account for optically thick regions ($\tau(r,\nu) = \Sigma(r) \kappa(\nu) > 1$)
by modifying Eq. (\ref{eq:flux}) as follows
\begin{equation}
F_{\nu}(r) dr= \left\{
\begin{array}{l l }
\frac{2k}{c^2D^2} \nu^2 \kappa(\nu) \Sigma(r) T(r) 2\pi r dr & \tau \leq 1  \\
\frac{2k}{c^2D^2} \nu^2  \frac{T(r)}{\tau^{1/4}} 2\pi r dr & \tau > 1.
\end{array} \right.
\label{eq:modflux}
\end{equation}
To calculate the integrated flux at $870 \mu$m from our simulated discs, we simply integrate Eq. (\ref{eq:modflux}) 
from the inner disc edge ($r = 1$ au) to the outer disc radius, using
$\kappa_o = 0.035$ cm$^2$ g$^{-1}$ and $\beta=1$.

\begin{figure*}
\begin{center}$
\begin{array}{cc}
\includegraphics[scale = 0.4]{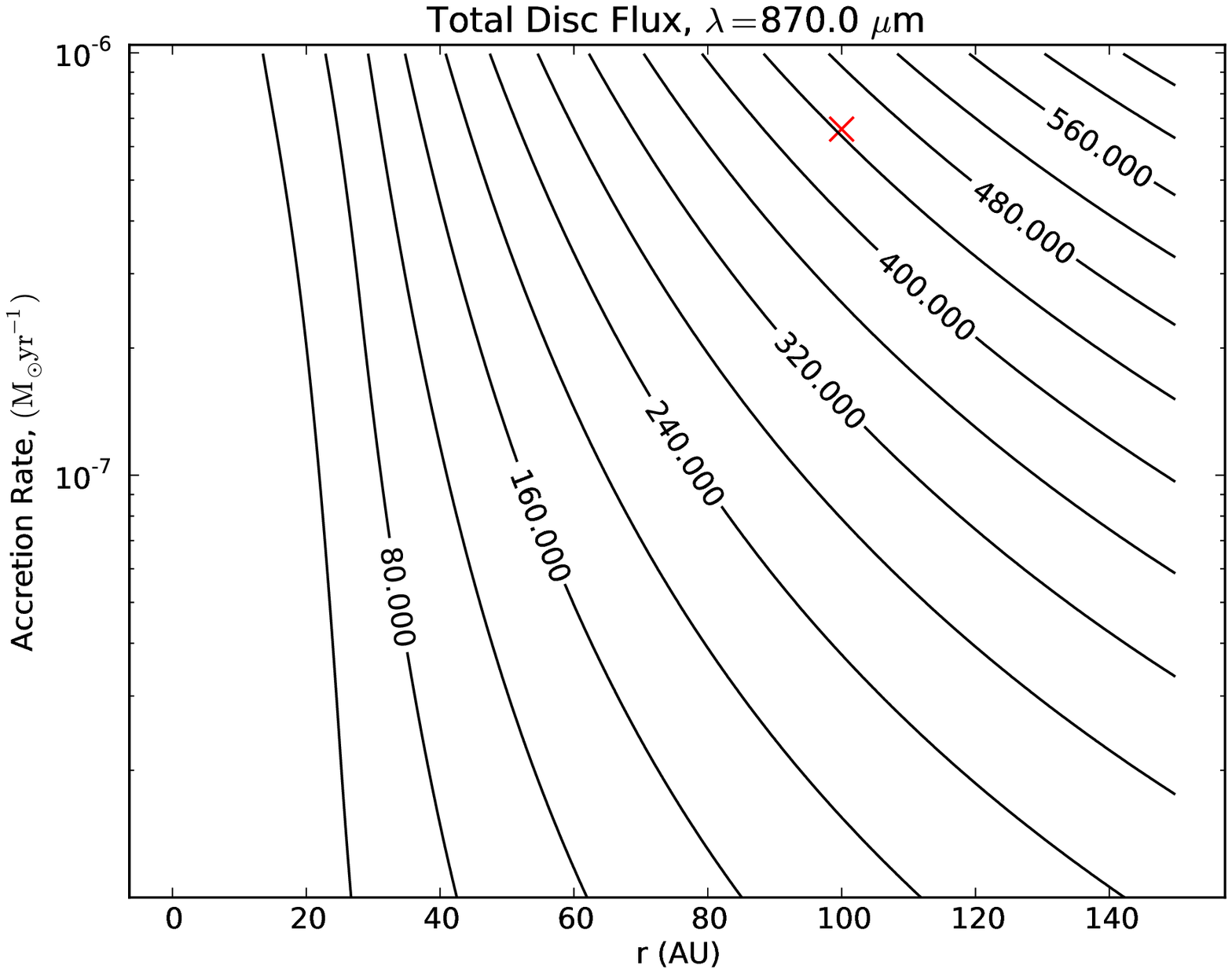} &
\includegraphics[scale=0.4]{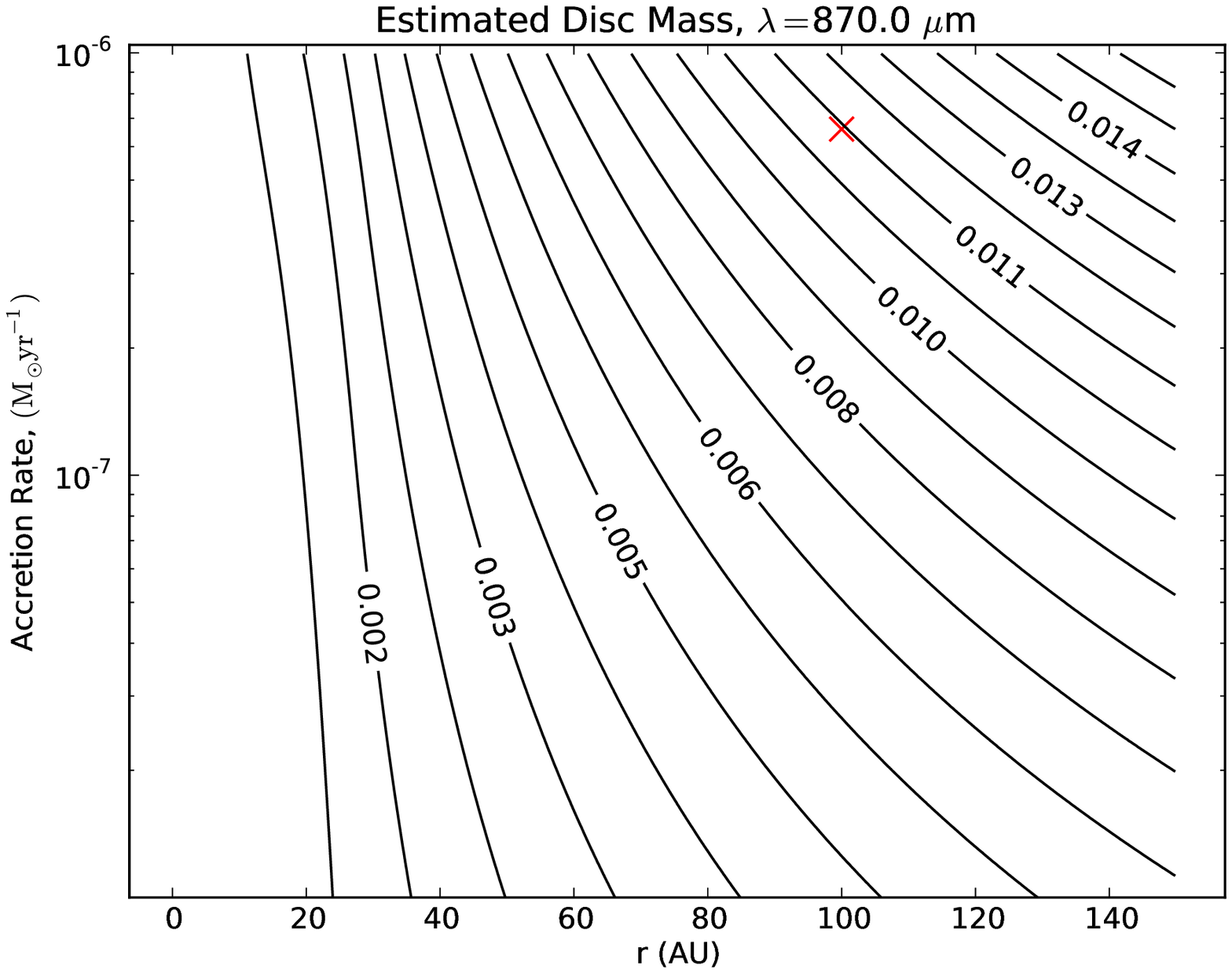} \\
\includegraphics[scale = 0.4]{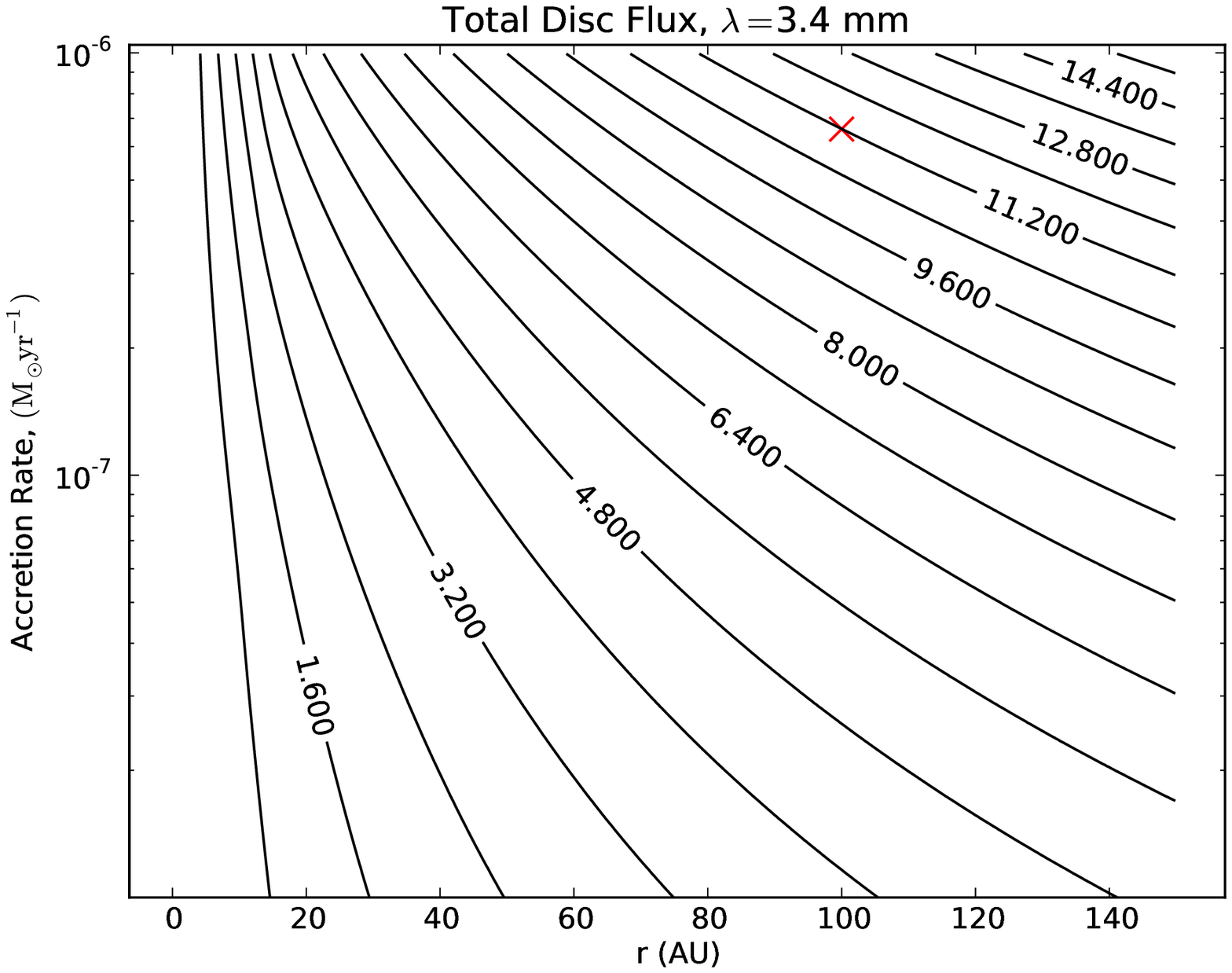} &
\includegraphics[scale=0.4]{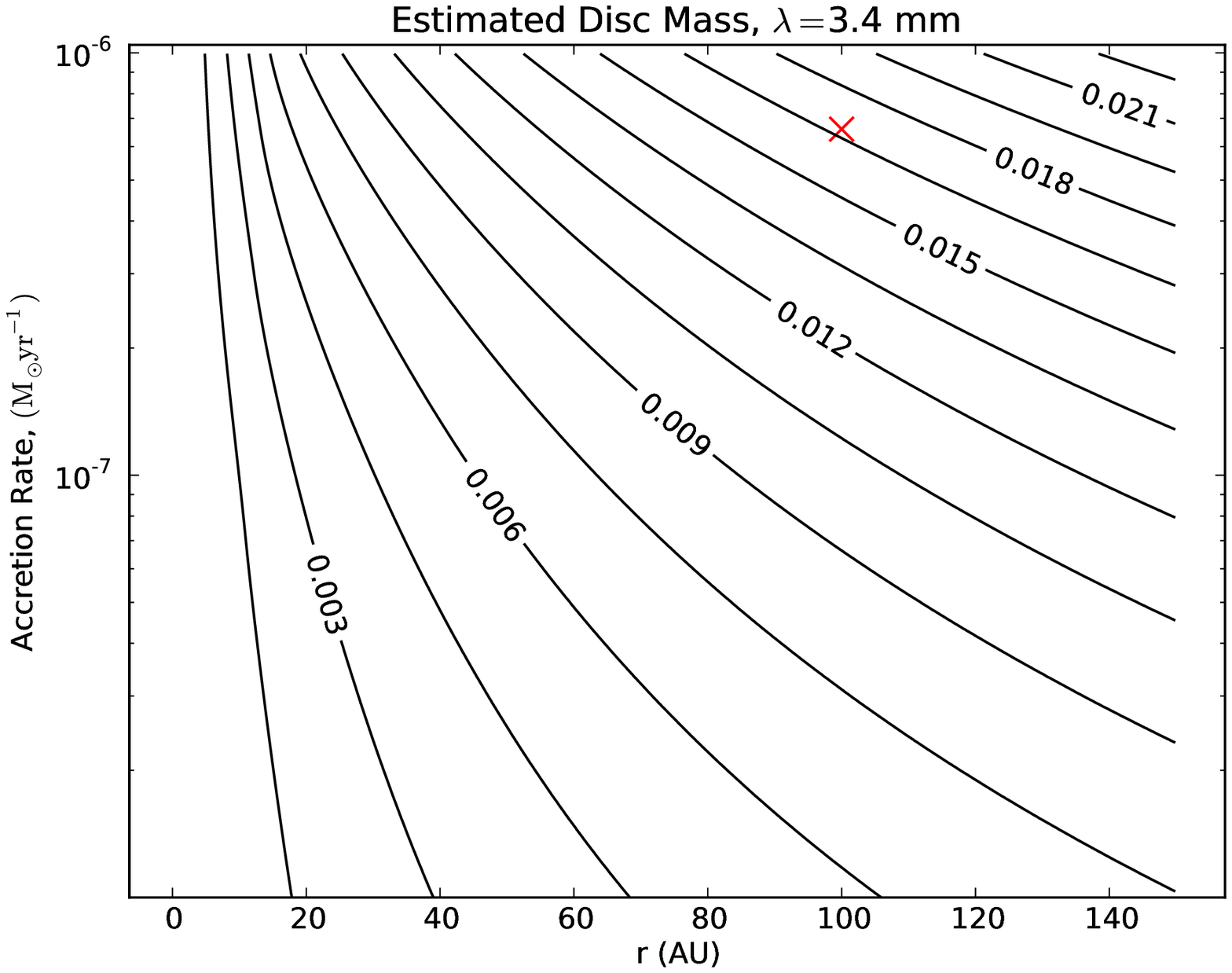} \\
\end{array}$
\caption{2D contours of the self-gravitating disc models observable properties, as a function of the steady-state accretion rate $\dot{M}$ and the disc's outer
  radius $r_{\rm out}$, for a star of mass $0.2 \msun$, assuming no external irradiation. Top left: the 870 $\mu$m flux (in mJy);  top right, the disc mass (in $\msun$) estimated from the 870 $\mu $m flux; bottom left, the 3.4 mm flux (in mJy);  bottom right, the disc mass (in $\msun$) estimated from the 3.4 mm flux.  Assuming an outer radius of $100$ au, the red crosses indicate what would be expected for a disc with an accretion rate similar to that of L1527 IRS.\label{fig:noirr}} 
\end{center}
\end{figure*}

The top left panel of Figure \ref{fig:noirr} shows contours of $870 \mu$m flux in mJy plotted agains $\dot{M}$ and outer disc
radius, $r_{\rm out}$.  For $r_{\rm out} = 100$ au and for $\dot{M} = 6.6 \times 10^{-7}$ M$_\odot$ yr$^{-1}$ the
$870 \mu$m flux is $\sim 475$ mJy. For $r_{\rm out} = 150$ au, this increases to $650$ mJy. 
The observed $870 \mu$m flux for L1527 IRS is $213.6 \pm 8.1$ mJy.  Despite our 
required disc masses (shown in Figure \ref{fig:mdot_r_q}) being an order of magnitude - or more -  
greater than that determined for L1527 IRS, the $870 \mu$m flux 
is only a factor of 2 to 3 greater than that observed.  Our calculation is also fairly simple and essentially
assumes a face-on geometry. The disc in L1527 IRS is edge-on and one might expect a reduction in flux at
these wavelengths by as much as a factor of $\sim 2$ due to this edge-on geometry (e.g., \citealt{whitney03}). 

Typically the disc mass is inferred from the long-wavelength flux using
\begin{equation}
M_{\rm disc} = \frac{D^2 F_\nu c^2}{2 \kappa(\nu) k \nu^2 T_{\rm dust}}.
\label{eq:Mdisc}
\end{equation}
Although our models have self-consistent disc temperatures, to compare with the results of \citet{tobin12}, we
use their value of $T_{\rm dust} = 30$ K.  The top right panel of Figure \ref{fig:noirr} shows the disc mass that would be inferred
from our simulated $870 \mu$m flux.  As expected, for $r_{\rm out} = 100$ au and $\dot{M} = 6.6 \times 10^{-7}$ M$_\odot$ yr$^{-1}$, 
the $870 \mu$m flux would suggest a disc mass of $\sim 0.011$ M$_\odot$, almost 10 times lower than the actual disc mass. 
Admittedly, if one compares the top left and right panels of Figures \ref{fig:noirr} one will notice that our estimate 
of the disc mass for a flux of $213$ mJy is slightly less than that determined by \citet{tobin12} ($0.00525$ M$_\odot$
rather than $0.007$ M$_\odot$), but this doesn't really change our main result.  Using the $870 \mu$m flux, 
the mass one would estimate for a self-gravitating disc accreting at $6.6 \times 10^{-7}$ M$_\odot$ yr$^{-1}$ around a star
of $M_* = 0.2$ M$_\odot$ would be $5 - 10$ times lower than the actual disc mass and only $2 - 3$ times higher than that
estimated for the disc around L1527 IRS.

\citet{tobin12} also measure 3.4 mm fluxes which they find to be $16.0 \pm 1.4$ mJy. Assuming $\beta = 1$ in Eq. (\ref{eq:opacity})
we can calculate the 3.4 mm fluxes from our disc models.  This is shown in the bottom left panel of  Figure \ref{fig:noirr}.  For a disc of
radius $r_{\rm out} = 100$ au and for $\dot{M} = 6.6 \times 10^{-7}$ M$_\odot$ yr$^{-1}$ our model predicts a $3.4$ mm flux of
$\sim 11.1$ mJy.  This is lower than, but similar to, that determined for L1527 IRS.

We can again estimate the disc mass using Eq. (\ref{eq:Mdisc}) which is shown in the bottom right panel of Figure \ref{fig:noirr}. Our simulated 
3.4 mm flux would suggest a disc mass (for properties similar to the of L1527 IRS) of $0.018$ M$_\odot$.  This is higher than the 
$870 \mu$m flux estimate ($0.011$ M$_\odot$) indicating that, at longer wavelengths, the disc is less optically thick.  The difference
is, however, much smaller than that seen by \citet{tobin12}. Their disc mass estimated from the $3.4$ mm flux ($0.025 \pm 0.003$ M$_\odot$) is 
3.6 times greater than the $870 \mu$m estimate.  They have more confidence in the $870 \mu$m flux as it is less affected by
assumption of $\beta$ but do suggest that this could indicate the disc being optically thick at $870 \mu$m, but our results would 
suggest that this alone cannot explain the observed difference the $870 \mu$m and $3.4$ mm fluxes.  We have assumed $\beta=1$, reflecting the current uncertainty regarding grain growth, and the degeneracy between $\beta$ and the disc mass when interpreting observations \citep{andrews05}.

We therefore also considered what would happen if we varied $\beta$.  For $\beta = 0.6$ we find that the $3.4$ mm flux 
(at $\dot{M} = 6.6 \times 10^{-7}$ M$_\odot$ yr$^{-1}$ and $r_{\rm out} = 100$ au) increases from $\sim 11$ mJy (when $\beta = 1$) to $\sim 16$ mJy.   
If one then assumes $\beta = 1$ when determining the disc mass from this flux, this increases the estimated disc mass from $0.018$ M$_\odot$ to
$0.026$ M$_\odot$.  The large difference in the measured $870 \mu$m and $3.4$ mm fluxes could, therefore, indicate that some grain growth has
taken place.  If so, this could also suggest that such discs would still have optically thick regions even at mm wavelengths.

\subsection{Irradiation at $T_{\rm irr} = 30$ K}

\noindent In the absence of external irradiation, the gravitational instability generates temperatures in the outer disc much lower than those estimated by \citet{tobin12}.  If we add external irradiation to ensure the outer disc cannot cool below 30 K, how does this affect the disc properties and the observable flux and mass?

\begin{figure}
\begin{center}
\psfig{figure=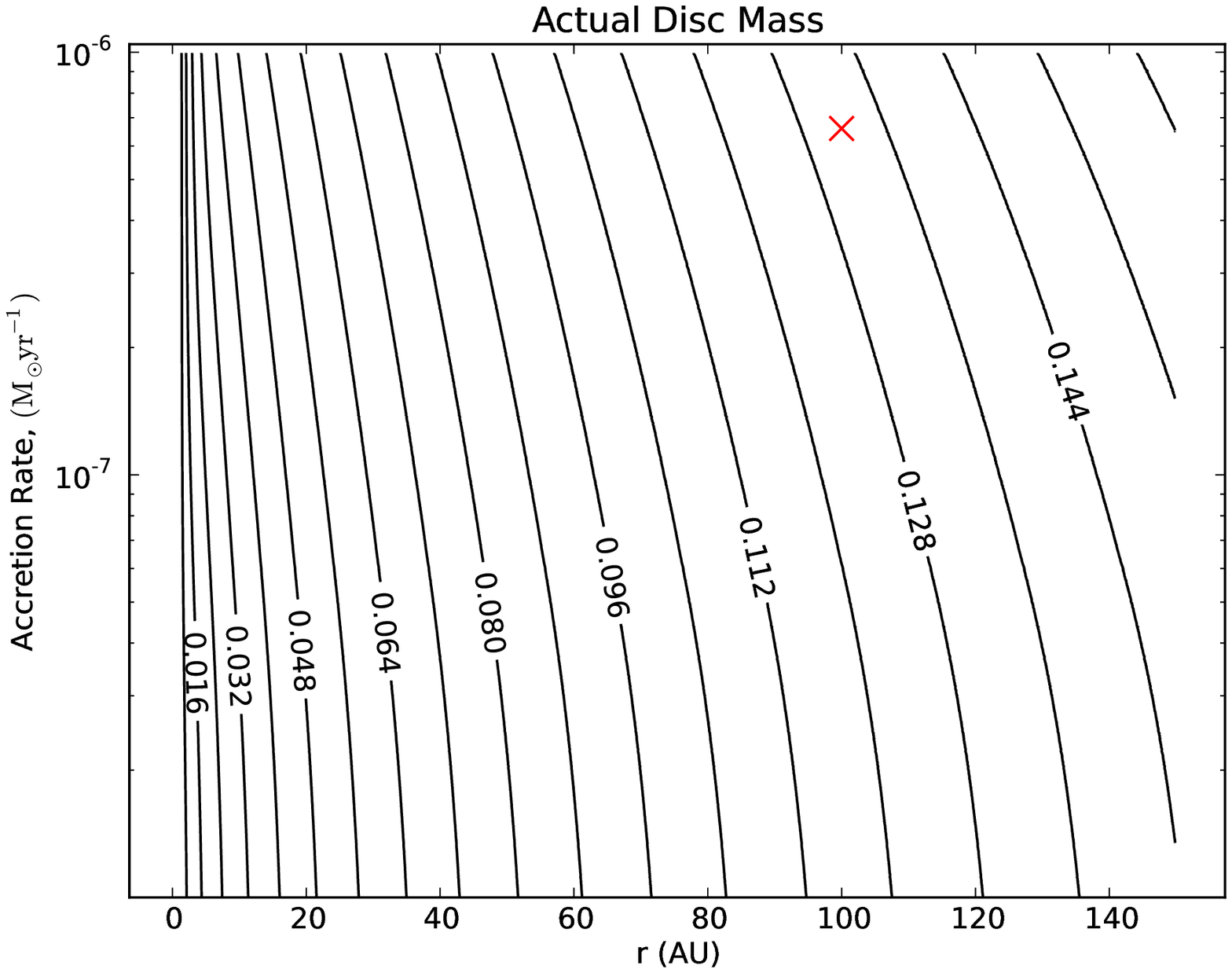,width=0.5\textwidth}
\caption{2D contours of the actual disc mass calculated from 
    self-gravitating disc models in an external irradiation field of $T_{\rm irr}=30$K, as a function of the
    steady state accretion rate $\dot{M}$, and the disc's outer radius
    $r_{\rm out}$, for a star of mass $0.2 \msun$.  The asterisk indicates
    the required disc mass, assuming an outer radius of 100 au, to produce
a mass accretion rate like that measured for L1527 IRS.}
\label{fig:mdot_r_q_30K}
\end{center}
\end{figure}

\noindent Figure \ref{fig:mdot_r_q_30K} shows the actual disc mass as a function of $\dot{M}$ and $r_{\rm out}$ in the presence of irradiation.  As irradiation tends to weaken the gravitational instability, the disc itself must be more massive for a given $\dot{M}-r$ locus to maintain a marginally unstable state \citep{cai08,kratter11,rice11,forgan13}.  This would push the actual disc mass of L1527 IRS to 0.132 $\msun$,  around two-thirds the mass of the central star.  The contours are almost vertical, indicating that the disc surface density profile is almost independent of accretion rate, and depends only on radius.  This is suggestive of heating due to the gravitational instability being superseded by the heating due to the irradiation, and the subsequent weakening of the instability, which occurs even at such large disc-to-star mass ratios \citep{rice09}.  As such, the disc surface density profile is no longer as centrally condensed as it was in the non-irradiated case. 

\begin{figure*}
\begin{center}$
\begin{array}{cc}
\includegraphics[scale = 0.4]{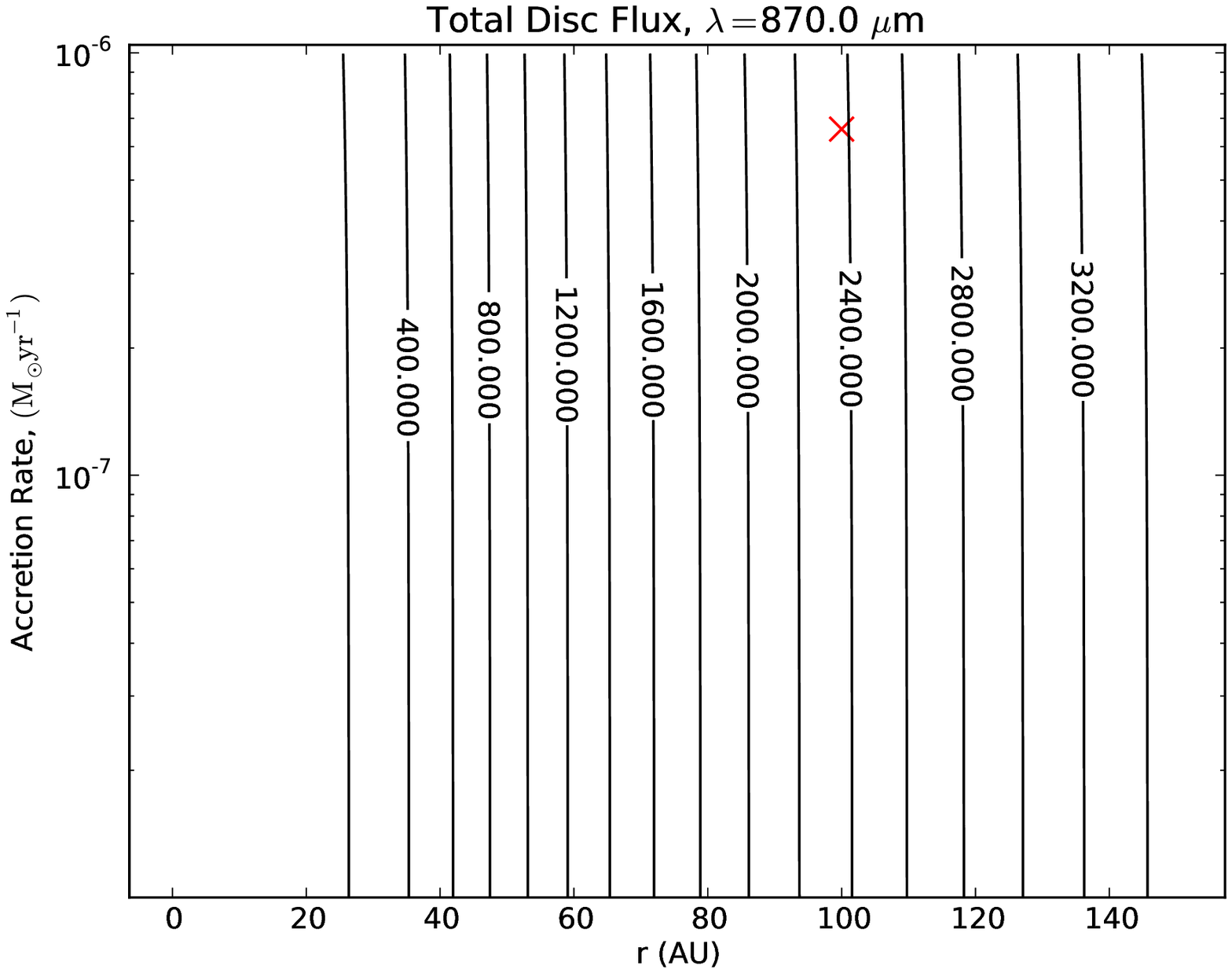} &
\includegraphics[scale=0.4]{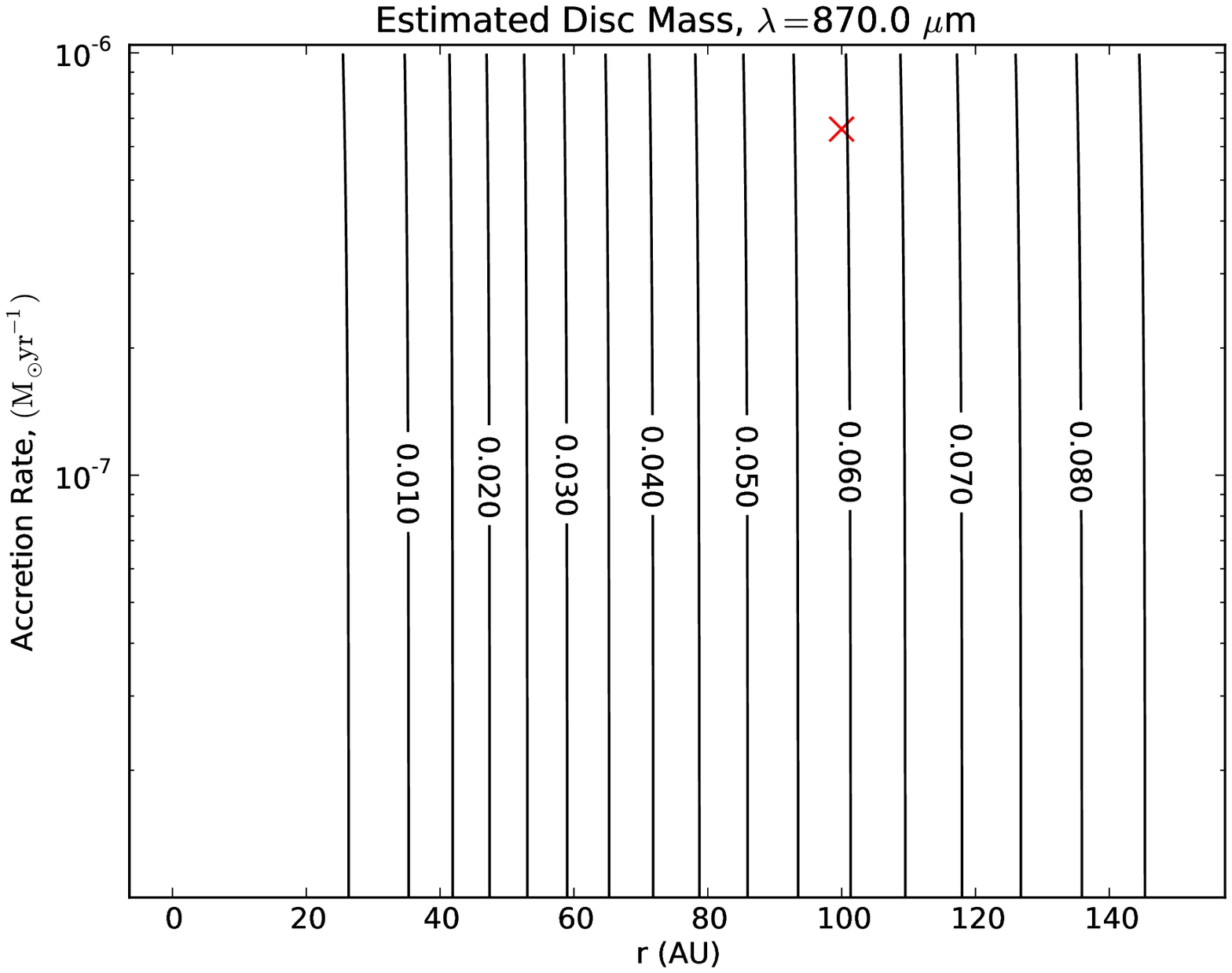} \\
\includegraphics[scale = 0.4]{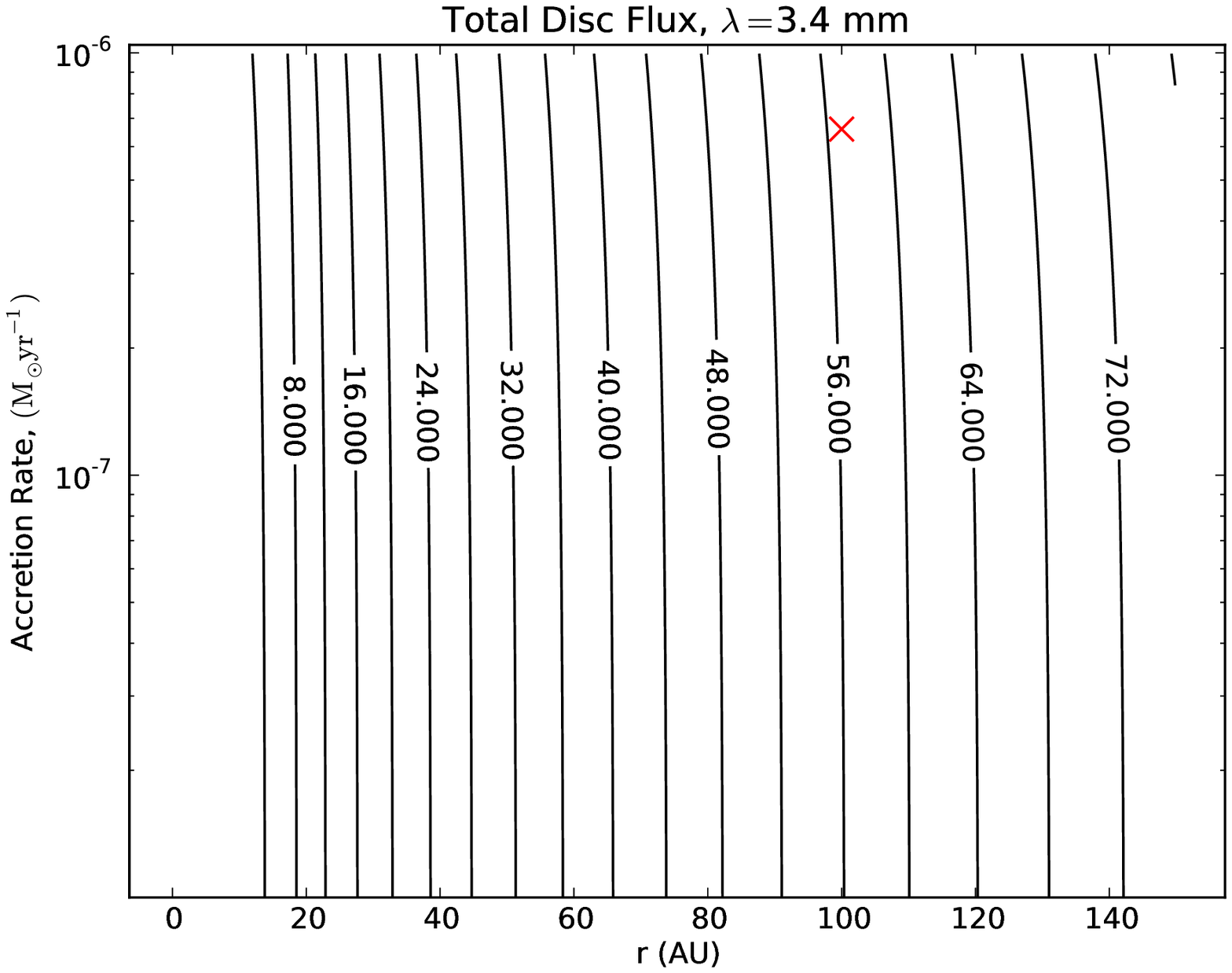} &
\includegraphics[scale=0.4]{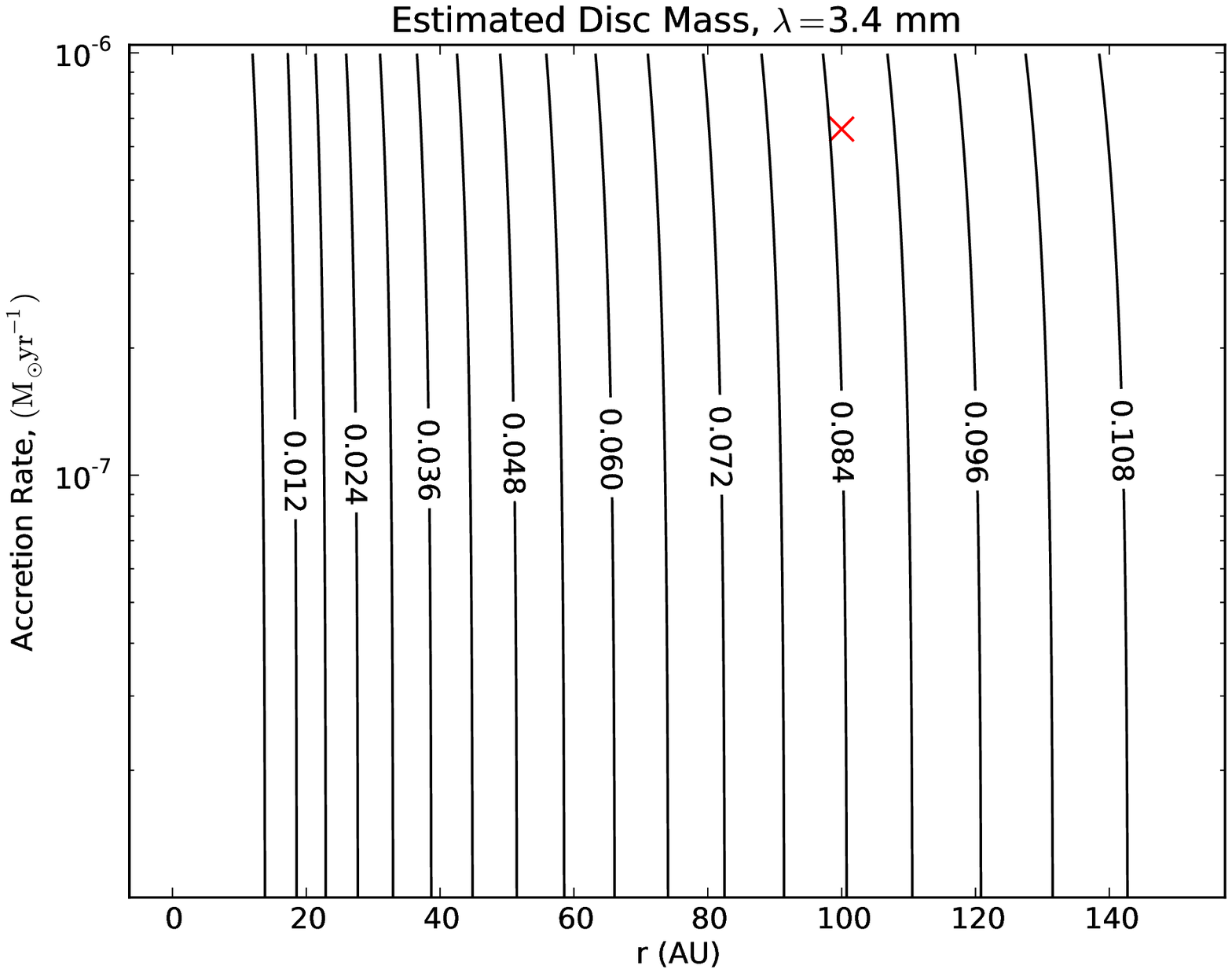} \\
\end{array}$
\caption{2D contours of the self-gravitating disc models observable properties, as a function of the steady-state accretion rate $\dot{M}$ and the disc's outer
  radius $r_{\rm out}$, for a star of mass $0.2 \msun$, assuming irradiation with a uniform temperature of $T_{\rm irr}  = 30$ K. Top left: the 870 $\mu$m flux (in mJy);  top right, the disc mass (in $\msun$) estimated from the 870 $\mu $m flux; bottom left, the 3.4 mm flux (in mJy);  bottom right, the disc mass (in $\msun$) estimated from the 3.4 mm flux.  Assuming an outer radius of $100$ au, the red crosses indicate what would be expected for a disc with an accretion rate similar to that of L1527 IRS.\label{fig:irr30K}} 
\end{center}
\end{figure*}

Adding this extra mass to the disc's outer regions boosts the observed flux by around a factor of 5 (left panels of Figure \ref{fig:irr30K}).  As the dust mass depends linearly on the flux, the estimated mass is also a factor of 5 larger (right panels of Figure \ref{fig:irr30K}).  

This is certainly less attuned with the observed flux than the non-irradiated case.  The uncertainty regarding the temperature profile of L1527 IRS leaves us uncertain as to the efficacy of self-gravitating disc models to explain the observations.  While it is likely that irradiation does set the temperature of the outer disc, modelling of other dust discs in Taurus in the sub-mm suggests that the disc temperature at 100 au could be up to a factor of 3 lower \citep{andrews05}. 

\subsection{Fragmentation}
Although we have primarily considered quasi-steady self-gravitating discs, it is thought that if such discs are sufficiently unstable they may undergo
fragmentation to form bound objects \citep{gammie01,rice03}.  This depends on how quickly such discs are able to cool and it is expected that it would
only be possible in the outer parts of such discs \citep{rafikov05,stamatellos07}.  The condition for fragmentation is that the effective gravitational viscosity
satisfies $\alpha > 0.06$ \citep{rice05}.  \citet{forgan11} suggest that once $\alpha$ exceeds this threshold, fragmentation also requires that the Jeans mass
changes sufficiently rapidly that a clump would actually form on a reasonably short timescale.  This allows one to estimate under what conditions
fragmentation should occur and, if fragmentation does occur, the resulting Jeans mass.  

In the non-irradiated case, our calculation show that L1527 IRS would need to maintain an accretion rate of around $\dot{M} \sim 8 \times 10^{-7} \mathrm{\msun \,yr^{-1}} $ or higher to fragment, and is therefore extremely close to fragmenting.  In the presence of irradiation, this critical accretion rate increases, and the disc is therefore stable \citep{forgan13}.  It is therefore probable that L1527 IRS will not fragment but if fragmentation has occurred,
it may be possible to detect enhanced emission consistent with an
object of a few Jupiter masses or larger forming via fragmentation in
such a disc \citep{greaves08}.


\section{Discussion and Conclusion}
The observation of a rotationally supported disc in the L1527 IRS protostellar system is the first confirmation
of a reasonably extended ($r_{\rm out} > 100$ au) disc around a class 0 protostar.  The disc mass estimated from 
the sub-mm flux, if correct, suggests that such a disc cannot be self-gravitating.  

We show here, that a self-gravitating disc can provide the observed mass accretion rate, but the disc mass would need to be at least an
order of magnitude greater than that observed.  Such discs are very centrally condensed \citep{clarke09,rice09} and so
the inner regions are optically thick even at long wavelengths.  Such discs may produce episodic outbursts\citep{vorobyov05,vorobyov10} and the observed FU Orionis outbursts \citep{hartmann96} may be driven by centrally condensed self-gravitating discs \citep{zhu09,zhu10}. 

Using reasonable values for the frequency dependent opacity, we
show that the fluxes produced by the model are similar to that observed for L1527 IRS if external irradiation is weak.  If these fluxes are then used to estimate the disc mass, these would give estimates $5 - 10$ times lower than the actual disc mass.  By varying the opacity spectral index $\beta$ from its canonical value of 1 down to 0.6, the fluxes obtained from the model match the observed values very closely.  This might suggest that significant grain growth is already underway in the Class 0 phase, perhaps due to the self-gravitating disc turbulence reducing the relative grain velocities allowing the grain growth process to be accelerated \citep{rice04,clarke09a}.  Such a hypothesis remains unconfirmable due to the degeneracies tying $\beta$ to the disc mass, not to mention the underlying uncertainties surrounding the grain composition and size distribution (\citealt{andrews05} and references therein).

However, in the presence of irradiation, the self-gravitating disc model produces fluxes up to 5 times higher than observed.  We therefore cannot claim that the observations of L1527 IRS show that the disc is massive and self-gravitating, especially as we do not model obscuration from the envelope in which the system is embedded.  If the envelope strongly obscures flux over a wide range of wavelengths, then the over-estimated flux produced by the disc models in the irradiated case may be greatly reduced, bringing them closer to the observed fluxes from L1527 IRS.    

Equally, given the extremely young age of the system and the difficulties of modelling the disc in the presence of irradiation, we can infer that the self-gravitating phase of protostellar discs is indeed constrained to the very earliest stages of star formation.  Despite possessing a high accretion rate, the self-gravitating disc model's over-estimate of flux in the presence of irradiation suggests that the disc may already have transitioned from a state in which self-gravity dominates as an angular momentum transport mode, to a state where MRI now dominates the transport.  If this is true, and the disc is still somewhat self-obscured, it is evident that self-gravity becomes ineffective in protostellar discs even while the disc-to-star mass ratio remains quite large \citep{rice09}.  This is illustrated by the disc mass becoming only weakly dependent on accretion rate in the presence of irradiation, i.e. they are not strongly centrally condensed, and are in fact more akin to the typical $\Sigma \propto r^{-1}$ powerlaw disc profiles  We suggest that the actual disc mass lies between the lower limit (established by observation) of $0.007 \pm 0.0007 \msun$ and the upper limit (established by this work) of $\sim 0.1 \msun$.


\section*{acknowledgements}
D.F. and K.R. acknowledge support from the Scottish Universities Physics Alliance (SUPA) 
and from the Science and Technology Facilities Council (STFC) through
grant ST/J001422/1. The authors would also like to thank John Tobin and Lee Hartmann for useful
discussions about the LR1527 IRS system, and the anonymous referee for their invaluable comments which greatly improved this paper.

\end{document}